# Smart Socket for Activity Monitoring

Manfred Sneps-Sneppe  
Ventspils University College  
Ventspils, Latvia  
manfreds.sneps@gmail.com

Dmitry Namiot  
Lomonosov Moscow State University  
Moscow, Russia  
dnamiot@gmail.com

*Abstract—* **In this short paper we consider the problem of monitoring physical activity in the smart house. The authors suggested a simple device that allows medical staff and relatives to monitor the activity for older adults living alone. This sensor monitors the switching-on of electrical devices. The fact of switching is seen as confirmation of physical activity. It is confirmed by SMS notifications to observers.**

## I. Introduction

The task of maintaining the quality of life for the elderly people is an urgent topic worldwide. One element of this quality is the ability to maximize long stay in a familiar, homely environment. However, independent living can be associated with several objective difficulties, require periodic assistance, as well as ongoing monitoring.

Accordingly, the task of monitoring (surveillance) for this type of patient is more than relevant for relatives, nursing services and medical facilities.

The base elements of the available monitoring systems are sensors technologies. For example, specialized sensors (or even mobile phones) equipped with an accelerometer and a gyroscope track physical activity [1]. It may be kinematic sensors [2], position sensors [3], etc.

A separate issue here is a set of sensors (and technologies) for remote obtaining of health indicators [4]. E-health (a telemedicine) is a separate and very rapidly developing area. Note, that the so-called wearable technologies will find here a very promising area for their applications [5].

Practical examples of use are often very complex and expensive systems. For example, the cost of installation of such a system is $5,500, with a monthly payment of $300 - $400 [6].

In this connection, we may mention a group of European programs. For example, Ambient Assisted Living joint programme (AAL) [7], SOPRANO-an ambient assisted living system to support older people at home [8]. A good overview of these programs is contained in papers [9,10].

AAL declares the following goals for the program:

New models of service delivery and care that contribute to greater self-reliance for older adults and greater support for informal careers;

Adapted living spaces that can improve the quality of their everyday lives;

New ways for older people to remain active, including contributing as volunteers or providing mutual support;

New ways of mobilizing active and trusted networks, both formal and informal, professional and in kind, to provide all types of support

Also, we can mention in this connection the domestic program on Internet of Things (IoT), proposed in [11], as well as the development, associated with home gateway [12].

## II. Smart socket

In this section we present our own device for activity monitoring. For the first time it was described in our paper [13].

The product's idea has been born from a rethinking of the well known device GSM Alarm - GSM socket (Figure 1)

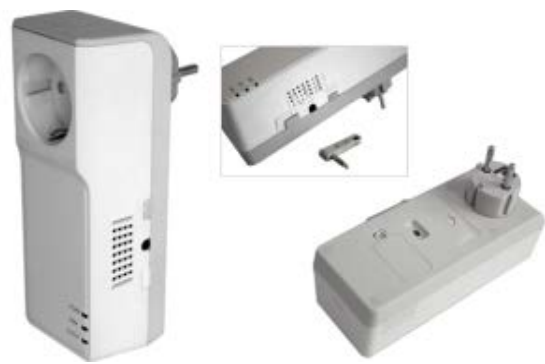

Fig. 1 GSM socket

These intelligent devices are designed to provide security and safety monitoring of your premises and control of electrical equipment remotely. The basic GSM power sockets are designed just for remote switching and reboot. Also they can send SMS about power failure alerts. And advanced solutions can do more. GSM socket could be a hub for sensors. They can send out temperature information via the mobile network, notify a mobile phone when someone breaks into home or in the case of fire or gas or water leak. They could have a built-in thermometer and function as a thermostat, which controls the turning on and off of equipment dependent on your settings. A built-in microphone allows you to hear remotely what is happening in the room. These devices are simple and what is even more important, they are very transparent for end-users.

Our idea is to reverse the operating principle. The proposed system does not control the power in the network. It should control the connection for external consumer. The idea is to control the switching-on of appliances via Smart Socket and send a SMS notification that some external (connected) device is turned on. The fact of switching is the confirmation for activity. The person, whom we watch, turns on some device. This socket (Figure 2) can be used, for example, to connect electric cookers (electric ignition for gas cookers). In this case, the switching of the cooker can be used as a confirmation of activity.

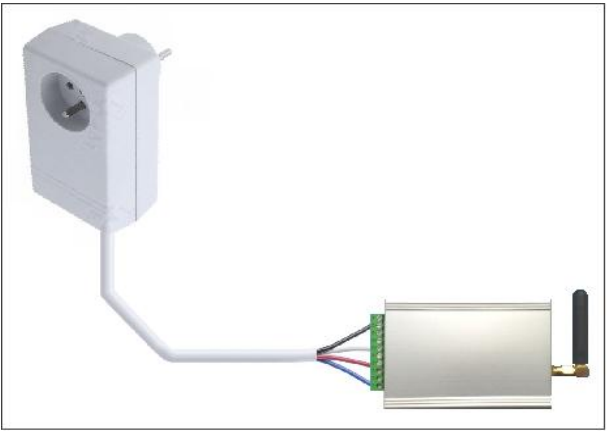

Fig.2 Smart socket

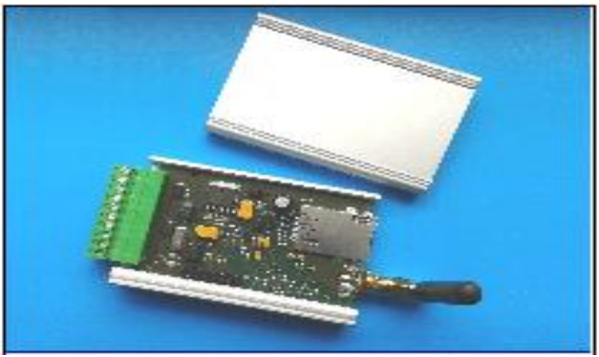

Fig. 3 GSM module

Technically, it is a current sensor with GSM modem. GSM-module (Figure 3) could be "programmed" via SMS. We must simply send one time SMS with phone number where to address SMS-notifications about connecting consumers. So, our Smart Socket will send SMS notifications about switching on electrical devices. There are two possible modes:

- directly send notifications to end-users (relatives, medical staff, etc);

- send notifications to intermediate server. And this server will route SMS messages to end-users.

The intermediate server (middleware) lets us estimate the behavior for individual sockets. For example, the fact of turning on of the particular device in the time frame from 10 till 11 in the morning for several consecutive days may be recorded as a pattern of behavior for that particular socket. In the presence of patterns, the service center (middleware) will be able to proactive send an alarm message to the destination number (family, social service, etc.) just due to absence socket's notifications in "usual" time. So, the fact that user does not turn on the electrical device could be used as a reason for alarms too. And, of course, the middleware server, having a full network connection, can send notifications using other transport mechanisms. For example, the middleware server cans notify end-users via push-notification [14] and Twitter [15].